\documentclass[sigconf]{acmart}

\usepackage{graphicx}

\usepackage{bm}
\usepackage{setspace}
\usepackage{amsmath,amsfonts,amsthm}
\usepackage{booktabs}

\usepackage{textcomp}
\usepackage{xcolor}
\usepackage[font=footnotesize]{subfig}
\usepackage{multirow}
\usepackage{float}

\setcopyright{acmcopyright}

\acmConference[SIGIR'20]{International ACM SIGIR Conference on Research and Development in Information Retrieval}{July 25-30, 2020}{Xi'an, China}
\acmYear{2020}
\copyrightyear{2020}

\author[F. Yu, Y. Zhu, Q. Liu, S. Wu, L. Wang, T. Tan]{Feng Yu$^{1,2,*}$, Yanqiao Zhu$^{1,2,*}$, Qiang Liu$^{3,4}$, Shu Wu$^{1,2,\dagger}$, Liang Wang$^{1,2}$, and Tieniu Tan$^{1,2}$}

\makeatletter
\def\authornotetext#1{
	\g@addto@macro\@authornotes{%
	\stepcounter{footnote}\footnotetext{#1}}%
}
\makeatother

\authornotetext{The first two authors contributed equally to this work.}
\authornotetext{To whom correspondence should be addressed.}

\affiliation{%
	\institution{$^1$Center for Research on Intelligent Perception and Computing, Institute of Automation, Chinese Academy of Sciences}
	\institution{$^2$School of Artificial Intelligence, University of Chinese Academy of Sciences}
	\institution{$^3$RealAI \qquad $^4$Tsinghua University}
}

\email{{feng.yu,shu.wu,wangliang,tnt}@nlpr.ia.ac.cn, yanqiao.zhu@cripac.ia.ac.cn, qiang.liu@realai.ai}

\def\authors{Feng Yu, Yanqiao Zhu, Qiang Liu, Shu Wu, Liang Wang, Tieniu Tan}

\begin{document}

\title{TAGNN: Target Attentive Graph Neural Networks for Session-based Recommendation}

\begin{abstract}
Session-based recommendation nowadays plays a vital role in many websites, which aims to predict users' actions based on anonymous sessions. There have emerged many studies that model a session as a sequence or a graph via investigating temporal transitions of items in a session. However, these methods compress a session into one fixed representation vector without considering the target items to be predicted. The fixed vector will restrict the representation ability of the recommender model, considering the diversity of target items and users' interests. In this paper, we propose a novel target attentive graph neural network (TAGNN) model for session-based recommendation. In TAGNN, target-aware attention adaptively activates different user interests with respect to varied target items. The learned interest representation vector varies with different target items, greatly improving the expressiveness of the model. Moreover, TAGNN harnesses the power of graph neural networks to capture rich item transitions in sessions. Comprehensive experiments conducted on real-world datasets demonstrate its superiority over state-of-the-art methods.
\end{abstract}


\maketitle

\section{Introduction}

Recommender systems are one of the most successful applications in the fields of data mining and machine learning research.
%With the tremendous amount of user behavioral data widely available, recommender systems help users alleviate the problem of information overload and find information of interest in many real-world websites, such as e-commerce sites, music and video platforms, etc.
Most previous work studies approaches that personalize the recommendation according to constantly recorded user profiles. In many real-world applications, however, such long-term profiles, even the users' identities may not exist. As an emerging method aiming to solve this problem, session-based recommendation predicts the next action (e.g., which item to click) of a user, given her previous behaviors within the ongoing session.

Considering the high practical values of session-based recommendation, many approaches have been proposed so far. Traditionally, Markov-chain-based methods \cite{Rendle:2010is} predict the user's next action solely based on the previous action. Such a strong independence assumption suffers from noisy data and thus restricts its use in session-based recommendation scenarios. Recent trends in recommender systems have led to a proliferation of studies using deep neural network techniques. Models based on recurrent neural networks (RNNs) have achieved promising performance. For example, \citeauthor{Hidasi:2016uq} propose a RNN-based method GRU4Rec \cite{Hidasi:2016uq} to model short-term preferences with gated recurrent units (GRUs). Recently, NARM \cite{Li:2017hk} proposes two RNN-based subsystems to capture users' local and global preference respectively. Similar to NARM, STAMP \cite{Liu:2018er} extracts users' potential interests using a simple multilayer perception model and an attentive network.

Despite their effectiveness, we would argue that those methods are still in their infancy. Previous work highlights that complex user behavioral patterns are of great significance for session-based recommendation \cite{Li:2017hk,Liu:2018er}. However, these sequence-based methods only model sequential transitions between consecutive items, with complex transitions neglected. Take  repeated purchases as an example, which is one of the most prominent behaviors in e-shopping scenarios. Suppose a session for a user is \(s = v_1 \rightarrow v_2 \rightarrow v_1 \rightarrow v_3\). Then, it is hard for these sequence-based methods to capture such a to-and-fro relationship between items. Specifically, they will be confused about the relationship between item \(v_1\) and items \((v_2, v_3)\). In this paper, we propose to discover the complex transitional patterns underneath sessions through session graphs \cite{Wu:2019ke}. By modeling items in sessions as session graphs, this natural means of encoding the abundant temporal patterns within sessions produces more accurate representation for each item.

Moreover, candidate items are usually abundant and users' interests are usually diverse. Previous work \cite{Li:2017hk,Liu:2018er,Wu:2019ke} represents one session using one embedding vector. That fixed-size vector represents all interests of a single user, which cannot express diverse user interests, and thereby limits the expressiveness of the recommender model. As a brute-force solution, we may consider enlarging the dimension of that fixed-length vector, which will in turn expose the risk of overfitting and deteriorate the model performance. However, we observe that it is not necessary to embed all user interests into one vector when making prediction for a specific candidate item. For example, suppose that a customer has a historical session of (swimming suits, purse, milk, frying pan). If we want to recommend a handbag for her, we focus on her interests in the purse rather than the frying pan. That is to say, the interests of a user with rich behaviors can be \emph{specifically activated given a target item} \cite{Zhou:2018bb}.
In this paper, we refine the proposed graph-based model through a novel target attention module. We term the resulting model as {\underline T}arget {\underline A}ttentive {\underline G}raph {\underline N}eural {\underline N}etworks for session-based recommendation, TAGNN\footnote{Code available at \url{https://github.com/CRIPAC-DIG/TAGNN}} for brevity. The proposed target attention module aims to \emph{adaptively activate user interests by considering the relevance of historical behaviors given a target item}. By introducing a local target attentive unit, specific user interests related to a target item in the current session are activated, which will benefit downstream session representations as a result.

\begin{figure}
	\centering
	\includegraphics[width=0.9\columnwidth]{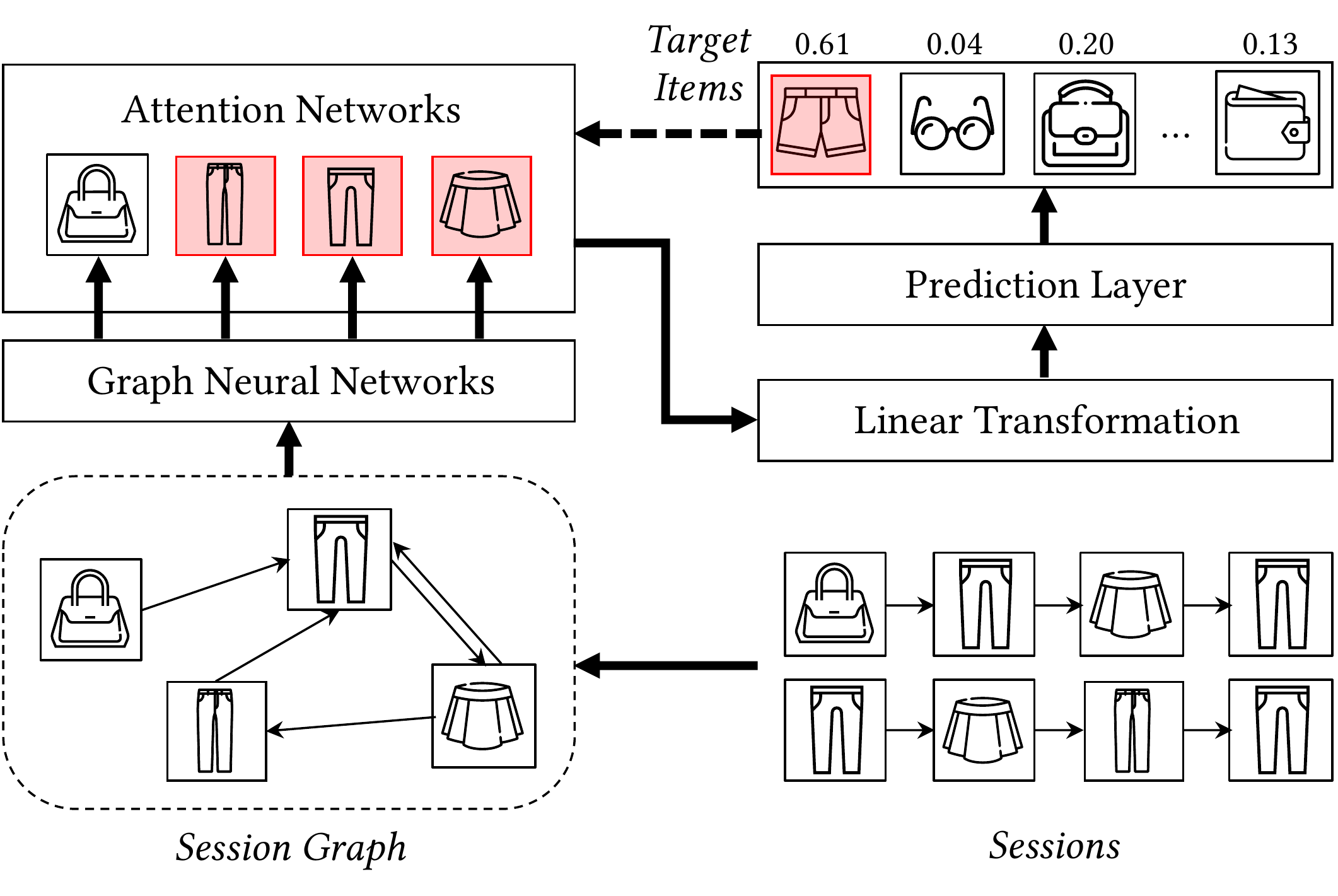}
	\caption{An overview of the proposed TAGNN method. Session graphs are constructed based on sessions at first. Then, graph neural networks capture rich item transitions in sessions. Last, from one session embedding vector, target-aware attention adaptively activates different user interests concerning varied target items to be predicted.}
	\label{fig:model}
\end{figure}

Figure \ref{fig:model} gives an overview of the TAGNN method. We first construct session graphs using items in historical sessions.
%It is worth mentioning that the way of modeling session graphs is flexible. Each session can either be modeled as a subgraph of the global session graph, or as a separate session graph.
After that, we obtain corresponding embeddings using graph neural networks to capture complex item transitions based on session graphs. Given the item embeddings, we employ a target-aware attentive network to activate specific user interests with respect to a target item. Following that, we construct session embeddings. At last, for each session, we can infer the user's next action based on item embeddings and the session embedding.
%Extensive experiments conducted on real-world datasets show that the proposed method consistently outperforms state-of-the-arts.

The main contribution of this work is threefold. Firstly, we model items in sessions as session graphs to capture complex item transitions within sessions. Then, we employ graph neural networks to obtain item embeddings. Secondly, to adaptively activate users' diverse interests in sessions, we propose a novel target attentive network. The proposed target attentive module can reveal the relevance of historical actions given a certain target item, which further improves session representations. Finally, we conduct extensive experiments on real-world datasets. The empirical studies show that our method achieves state-of-the-art performance.

\section{The Proposed Method: TAGNN}

\subsection{Problem Statement and Constructing Session Graphs}

In session-based recommendation, an anonymous session can be represented by a list \(s = [v_{s,i}]_{i=1}^{s_n}\) ordered by timestamps and we denote \(V = \{v_i\}_{i=1}^m\) as the set consisting of all unique items (e.g., user clicks) involved in sessions. Session-based recommendation aims to predict the next action \(v_{s,s_n+1}\) given session \(s\). Our model produces a ranking list of probabilities for all candidate items and items with top-\(k\) probability values will be selected for recommendation.

In our model, we represent each session \(s\) as a directed \emph{session graph} \(\mathcal{G}_s = (\mathcal{V}_s, \mathcal{E}_s, \bm{A}_s)\), where \(\mathcal{V}_s, \mathcal{E}_s, \bm{A}_s\) are the node set, the edge set, and the adjacency matrix, respectively. In this graph \(\mathcal{G}_s\), each node represents an item \(v_{s,i} \in V\) and each edge \((v_{s,i-1}, v_{s,i}) \in \mathcal{E}_s\) represents a user visits item \(v_{s,i-1}\) and \(v_{s,i}\) consecutively. Here we define \(\bm{A}_s\) as the concatenation of two adjacency matrices \(\bm{A}_s^{\text{(out)}}\) and \(\bm{A}_s^{\text{(in)}}\) to reflect the bidirectional relationship between items in sessions, where \(\bm{A}_s^{\text{(out)}}\) and \(\bm{A}_s^{\text{(in)}}\) represents weighted connections of outgoing and incoming edges respectively.
%For example, consider a session \(s = [v_1, v_2, v_3, v_1, v_3, v_4]\), we construct a session graph $\mathcal{G}_{s}$ as shown in Figure \ref{fig:session-graph}. Considering items may appear in the session repeatedly, we conduct row normalization for the two adjacency matrices.
%
%\begin{figure}
%	\centering
%	\includegraphics[width=\columnwidth]{figures/adjcency-matrix.pdf}
%	\caption{An example of a session graph \(\mathcal{G}_s\) and the corresponding adjacency matrix \(\bm{A}_s = [\bm{A}_s^{\text{(in)}}; \bm{A}_s^{\text{(out)}}]\).}
%	\label{fig:session-graph}
%\end{figure}
Here we use the same strategy of constructing session graphs as SR-GNN \cite{Wu:2019ke}. However, it is flexible to adopt different mechanisms of constructing the session graph for different session-based recommendation scenarios.
%For example, if the number of items in each session varies greatly, we may consider to model a global session graph containing all unique items. Under this setting, each session becomes a subgraph. Moreover, in a more generalized setting, if nodes contain meta-information such as tags and descriptive texts, we may concatenate the connection matrix $\bm{A}_s$ with the node feature matrix.
For clarity, we omit subscript \(s\) for referring item embeddings hereafter.

\subsection{Learning Item Embeddings}

After constructed session graphs, we transform every node \(v_i \in V\) into a unified embedding space. The resulting vector \(\bm{v}_i \in \mathbb{R}^d\) is a \(d\)-dimensional representation of item \(v_i\) obtained using graph neural networks. Then, we can represent each session \(s\) using item embeddings. The graph neural network (GNN) \cite{Scarselli:2009ku,Kipf:2016tc,Li:2016vj} is a class of widely used deep learning models. GNNs generate node representations on top of graph topology, which models complex item connections. Therefore, they are particularly suitable for session-based recommendation. In this paper, we employ gated graph neural networks (GGNNs) \cite{Li:2016vj}, a variant of GNN, to learn the node vectors. Formally, for node $v_{s,i}$ of graph $\mathcal{G}_s$, its update rules are:
\begin{align}
	\bm{a}^{(t)}_{s,i} & = \bm{A}_{s,i:} \left[\bm{v}^{(t-1)}_1, \dots ,\bm{v}^{(t-1)}_{s_n}\right]^\top \bm{H} + \bm{b}, \label{eq:node-representation}\\
	\bm{z}^{(t)}_{s,i} & = \sigma\left(\bm{W}_z\bm{a}^{(t)}_{s,i}+\bm{U}_z\bm{v}^{(t-1)}_{i}\right), \label{eq:update-gate}\\
	\bm{r}^{(t)}_{s,i} & = \sigma\left(\bm{W}_r\bm{a}^{(t)}_{s,i}+\bm{U}_r\bm{v}^{(t-1)}_{i}\right), \label{eq:reset-gate}\\
	\widetilde{\bm{v}^{(t)}_{i}} & = \tanh\left(\bm{W}_o \bm{a}^{(t)}_{s,i}+\bm{U}_o \left(\bm{r}^{(t)}_{s,i} \odot \bm{v}^{(t-1)}_{i}\right)\right), \label{eq:candidate-state}\\
	\bm{v}^{(t)}_{i} & = \left(1-\bm{z}^{(t)}_{s,i} \right) \odot \bm{v}^{(t-1)}_{i} + \bm{z}^{(t)}_{s,i} \odot \widetilde{\bm{v}^{(t)}_{i}} , \label{eq:final-state}
\end{align}
where \(t\) is the training step, \(\bm{A}_{s,i:} \in \mathbb{R}^{1 \times 2n}\) is the \(i\)-th row in matrix $\bm{A}_s$ corresponding to node $v_{s,i}$, \(\bm{H} \in \mathbb{R}^{d \times 2d}\) and \(\bm{b} \in \mathbb{R}^d\) are weight and bias parameter respectively, \(\left[\bm{v}^{(t-1)}_1, \dots , \bm{v}^{(t-1)}_{s_n}\right]\) is the list of node vectors in session $s$, \(\bm{z}_{s,i} \in \mathbb{R}^{d \times d}\) and \(\bm{r}_{s,i} \in \mathbb{R}^{d \times d}\) are the reset and update gates respectively, \(\sigma(\cdot)\) is the sigmoid function, and \(\odot\) denotes element-wise multiplication. For each session graph $\mathcal{G}_s$, the GGNN model propagates information between neighboring nodes. The update and reset gate decides what information to be preserved and discarded respectively. 

\subsection{Constructing Target-Aware Embeddings}

Previous work captures users' interests only using intra-session item representations. In our model, once we obtained the node vector for each item, we proceed to construct target embeddings, to adaptively consider the relevance of historical behaviors concerning target items. Here we define the target items as all candidate items to predict. Usually, a user's action given a recommended item only matches a part of her interests. To model this process, we design a novel target attention mechanism to calculate soft attention scores over all items in the session with respect to each target item.

In this section, we introduce a local target attentive module to calculate attention scores between all items \(v_i\) in session \(s\) and each target item \(v_t \in V\). Firstly, a shared non-linear transformation parameterized by a weight matrix \(\bm{W} \in \mathbb{R}^{d \times d}\) is applied to every node-target pair. Then, we normalize the self-attention scores using the softmax function:
\begin{equation}
	\beta_{i,t} = \operatorname{softmax}(e_{i,t}) = \frac{\exp \left(\bm{v}_t^\top \bm{W} \bm{v}_{i} \right) } {\sum_{j=1}^{m} \exp \left(\bm{v}_t^\top \bm{W} \bm{v}_{j}\right)}.
\end{equation}
Finally, for each session \(s\), the users' interests towards a target item \(v_t\) is represented by \(\bm{s}^t_\text{target} \in \mathbb{R}^d\), as given below:
\begin{equation}
	\bm{s}_\text{target}^t = \sum_{i=1}^{s_n} \beta_{i,t} \bm{v}_i.
\end{equation}
The obtained \emph{target embedding} for representing users' interests varies with different target items.

\subsection{Generating Session Embeddings}

In this section, we further exploit users' short- and long-term preference exhibited in the current session \(s\) using node representations involved in session \(s\). The resulting two representations along with the user's target embedding will be further concatenated to generate better session embeddings.

\textbf{Local embedding.}
As the user's final action is usually determined by her last action, we simply represent the user's short-term preference as a \emph{local embedding} \(\bm{s}_\text{local} \in \mathbb{R}^d\) as the representation of the last-visited item \(v_{s,s_n}\).

\textbf{Global embedding.}
Then, we represent the user's long-term preference as a \emph{global embedding} $\bm{s}_\text{global} \in \mathbb{R}^d$ by aggregating all involved node vectors. We adopt another soft-attention mechanism to draw dependencies between the last-visited item and each item involved in the session:
\begin{align}
	\alpha_i &= \bm{q}^\top \, \sigma(\bm{W}_1 \bm{v}_{s_n} + \bm{W}_2 \bm{v}_{i} + \bm{c}), \\
	\bm{s}_\text{global} & = \sum\limits_{i = 1}^{s_n} {\alpha_i \bm{v}_{i}},
\end{align}
where \(\bm{q}, \bm{c} \in \mathbb{R}^d\) and \(\bm{W}_1, \bm{W}_2 \in \mathbb{R}^{d \times d}\) are weight parameters.

\textbf{Session embedding.}
Finally, we generate the session embedding $\bm{s}$ of session \(s\) by taking linear transformation over the concatenation of the local and global embeddings and the target embedding:
\begin{equation}
	\bm{s}_t = \bm{W}_3 [\bm{s}^t_\text{target}; \bm{s}_\text{local} ; \bm{s}_\text{global}],
\end{equation}
where \(\bm{W}_3 \in \mathbb{R}^{d \times 3d}\) projects the three vectors into one embedding space \(\mathbb{R}^d\). Please kindly note that we generate different session embeddings for each target item.

\subsection{Making Recommendation}
After obtaining all item embeddings and session embeddings, we compute the recommendation score \(\hat{z_t}\) for each target item \(v_t \in V\) by taking inner-product of item embedding \(\bm{v}_t\) and session representation \(\bm{s}\). Following that, we use the softmax function over all unnormalized scores \(\bm{z}\) for all target items and get the final output vector:
\begin{align}
	\hat{z_t} & = \bm{s}_t^\top \bm{v}_t, \\
	\hat{\bm{y}} & = \operatorname{softmax}\left( \hat{\bm{z}} \right).
\end{align}
Here \(\hat{\bm{y}} \in \mathbb{R}^m\) denotes the probabilities of nodes being the next action in \(s\). The items with the top-\(k\) probabilities in \(\hat{\bm{y}}\) will be selected as recommended items.

For training the model, we define the loss function as the cross-entropy of the prediction and the ground truth:
\begin{equation}
	\mathcal{L}(\hat{\bm{y}}) = -\sum_{i = 1}^{m} \bm{y}_i \log{(\hat{\bm{y}_i})} + (1 - \bm{y}_i) \log{(1 - \hat{\bm{y}_i})},
\end{equation}
where $\bm{y}$ denotes the one-hot encoding vector of the ground truth items. We use the back-propagation through time (BPTT) algorithm to train the proposed model.

\section{Experiments}

In this section, we aim to answer the following two questions:

\textbf{RQ1.}
Does the proposed TAGNN achieve state-of-the-art performance compared with existing representative baseline algorithms?

\textbf{RQ2.}
How do different schemes for representing user interests affect the model performance?

\subsection{Experimental Configurations}

\paragraph{Datasets.}
We evaluate the proposed method using two widely-used real-world datasets Yoochoose\footnote{http://2015.recsyschallenge.com/challenge.html} and Diginetica\footnote{http://cikm2016.cs.iupui.edu/cikm-cup}, obtained from two contests in data mining conference RecSys 2015 and CIKM 2016 respectively. 
For fair comparison, we closely follow the same data preprocessing scheme as \citet{Li:2017hk,Liu:2018er,Wu:2019ke}. Specifically, we drop items appearing less than 5 times and sessions consisting of less than 2 items. For generating training and test sets, sessions of last days are used as the test set for Yoochoose, and sessions of last weeks as the test set for Diginetica. For an existing session \(s = [v_{s,1}, v_{s,2}, \dots, v_{s,s_n}]\), we generate a series of input session sequences and corresponding labels as \(([v_{s,1}], v_{s,2}), ([v_{s,1}, v_{s,2}], v_{s,3}), \dots, ([v_{s,1}, v_{s,2}, \dots, v_{s,s_{n-1}}], v_{s,s_n})\).
%, where the list enclosed with square brackets \([v_{s,1}, v_{s,2}, \dots, v_{s,s_{n-1}}]\) is the input session and \(v_{s,s_n}\) is the label of that session.
Since the Yoochoose dataset is too large, we only use its the most recent 1/64 fractions of the training sessions, denoted as Yoochoose 1/64.
%, similar to \citet{Wu:2019ke}.
%The statistics of datasets used throughout experiments are summarized in Table \ref{tab:dataset-statistics}.
%
%\begin{table}
%	\centering
%	\caption{Statistics of datasets used in the experiments}
%	\resizebox{0.7\columnwidth}{!}{
%	\begin{tabular}{cccc}
%    	\toprule
%		Statistics & Yoochoose 1/64 & Diginetica \\
%		\midrule
%		\# Clicks & 557,248 & 982,961 \\
%		\# Training sessions & 369,859 & 719,470 \\
%		\# Test sessions & 55,898 & 60,858 \\
%		\# Unique items & 17,377 & 43,097 \\
%		Average length & 6.16 & 5.12 \\
%		\bottomrule
%	\end{tabular}
%	}
%	\label{tab:dataset-statistics}
%\end{table}

\paragraph{Baselines.}
To evaluate the performance of the proposed method, we comprehensively compare TAGNN with representative baselines. The traditional baselines include (a) frequency-based methods POP and S-POP,
% which simply generate most frequent items as recommendation, 
(b) similarity-based method Item-KNN \cite{Sarwar:2001kx},
% which recommends the most similar item to the previously visited item,
and (c) factorization-based methods Bayesian personalized ranking (BPR-MF) \cite{Rendle:2009wp}
%, which optimizes a pairwise ranking function, 
and factorizing personalized Markov chain model (FPMC) \cite{Rendle:2010is}.
%, which combines Markov chain with matrix factorization.
We also consider deep learning baselines, including RNN-based recommender model GRU4REC \cite{Hidasi:2016uq},
% which employs RNNs to model user sequences,
neural attentive recommender model (NARM) \cite{Li:2016vj},
% which designs two attention networks to capture user preference,
short-term attention/memory priority model (STAMP) \cite{Liu:2018er},
% which captures users' general interests of the current session and current interests of the last click.
and GNN-based recommender model SR-GNN \cite{Wu:2019ke}.

\paragraph{Evaluation Metrics.}
We adopt two commonly-used metrics for evaluation, including Precision@20 and MRR@20. The former one evaluates the proportion of correct recommendation in an unranked list, while the latter one further considers the position of correct recommended items in a ranked list.

\paragraph{Hyperparameter Setup.}
Following previous methods \cite{Liu:2018er,Wu:2019ke}, we set \(d=100\) for hidden dimensionality in all experiments. We tune other hyperparameters based on a random 10\% validation set. The initial learning rate for Adam is set to 0.001 and will decay by 0.1 after every 3 training epochs. The batch size is set to 100 for both datasets and the \(\ell_2\) penalty is set to \(10^{-5}\).

\subsection{Comparison with Baseline Methods (RQ1)}
To evaluate the performance of the proposed method, we firstly compare it with existing representative baselines (RQ1). The overall performance in terms of Precision@20 and MRR@20 is summarized in Table \ref{tab:result-baseline-algorithms}, with the highest performance highlighted in boldface.

\begin{table}
	\centering
	\small
	\caption{The performance of TAGNN compared with other baseline methods using two datasets.}
	\resizebox{0.9\columnwidth}{!}{
	\begin{tabular}{ccccc}
	\toprule
	\multirow{2}[0]{*}{Method} & \multicolumn{2}{c}{Diginetica} & \multicolumn{2}{c}{Yoochoose 1/64} \\
	\cmidrule(lr){2-3} \cmidrule(lr){4-5}
		& Precision@20 & MRR@20 & Precision@20 & MRR@20 \\
	\midrule
	POP   & 0.89  & 0.20  & 6.71  & 1.65  \\
	S-POP & 21.06  & 13.68  & 30.44  & 18.35  \\
	Item-KNN & 35.75  & 11.57  & 51.60  & 21.81  \\
	BPR-MF & 5.24  & 1.98  & 31.31  & 12.08  \\
	FPMC  & 26.53  & 6.95  & 45.62  & 15.01  \\
	GRU4REC & 29.45  & 8.33  & 60.64  & 22.89  \\
	NARM  & 49.70  & 16.17  & 68.32  & 28.63  \\
	STAMP & 45.64  & 14.32  & 68.74  & 29.67  \\
	SR-GNN & 50.73  & 17.59  & 70.57  & 30.94  \\
	\cmidrule(lr){1-5}
	TAGNN & \textbf{51.31} & \textbf{18.03} & \textbf{71.02} & \textbf{31.12} \\
	Improv.(\%) & 1.14 & 2.50 & 0.64 & 0.58 \\
	\bottomrule
	\end{tabular}
	}
	\label{tab:result-baseline-algorithms}
\end{table}

In all, TAGNN aggregates session items into session graphs and further considers modeling user preference through target-aware attention. It is apparent from the table that the proposed TAGNN model achieves state-of-the-art performance on all datasets in terms of Precision@20 and MRR@20, which confirms the effectiveness of the proposed method.

This table is quite revealing in several ways. Firstly, traditional methods including POP and S-POP achieve poor performance. They mainly emphasize items with high co-occurrence, which is over-simplified in session-based recommendation. Interestingly, the simple method Item-KNN still shows favorable performance, compared with POP, BPR-MF, and FPMC. Without knowing the sequential information, Item-KNN only recommends items with high similarity. This may be explained by the fact that latent factors representing user preference play a key role in generating accurate recommendation. It can also be seen that Item-KNN surpasses most Markov-chain-based methods, such as BPR-MF and FPMC, which demonstrates that modeling limited dependencies in session sequence is not realistic in session-based recommendation scenarios.

Secondly, there is a clear trend that deep learning methods greatly outperform conventional models. These methods have a stronger capability to capture complex user behavior, leading to superior performance over traditional ones. Sequential models such as GRU4REC and NARM only considers single-way transitions between successive item. Compared with SR-GNN, which further models session as graphs and is able to capture more implicit connections between user clicks, these methods neglect complex item transitional patterns. However, the performance of these models is still inferior to that of the proposed method. TAGNN elaborates graph-based models by further considering user interests with target-aware attentions. This mechanism specifically activates diverse user interests given different target items, which improves the expressiveness of the recommender model. In summary, these results demonstrate the efficacy of the proposed TAGNN method.

\subsection{Ablation Studies (RQ2)}

\begin{figure}
	\centering
	\includegraphics[width=0.9\columnwidth]{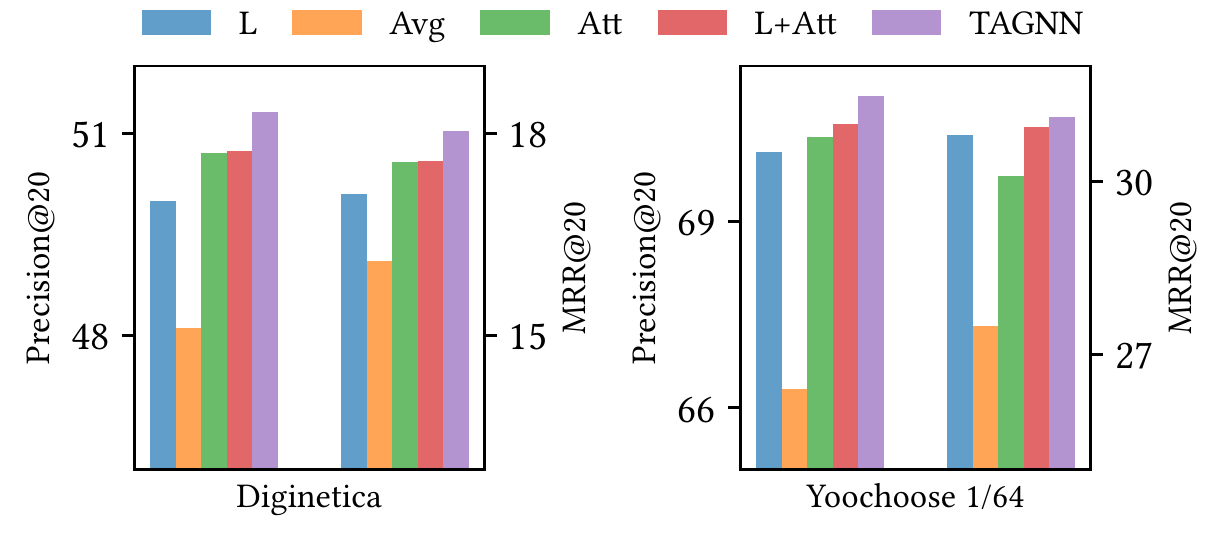}
	\caption{The performance of different session representations.}
	\label{fig:ablation-study}
\end{figure}

We conduct ablation studies on session embedding strategies in this section. We design 4 model variants to analyze how different representations for user preference affect the model performance (RQ2): (a) local embedding only (TAGNN-L), (b) global embedding using average pooling only (TAGNN-Avg), (c) attentive global embedding (TAGNN-Att), and (d) local embedding plus attentive global embedding (TAGNN-L+Att). Figure \ref{fig:ablation-study} presents experimental results using different session representations.

It is apparent from this figure that the hybrid embedding strategy used by TAGNN achieves the best performance on all datasets, which verifies the necessity of explicitly incorporating target-aware attention for better representing user interests. Also, it is clear that the mixed use of attentive global embedding with local embedding outperforms other opponents, demonstrating the necessity to capture both long- and short-term preference exhibited within sessions. Please note that the performance of TAGNN-L, which merely uses the last item as the session representation stands out in this figure. It indicates that the last item has a great impact on a user's final action. Moreover, we can observe that using average pooling over items in session achieves bad performance. This phenomenon may be explained by the diversity of user behavior in the session, which further highlights the importance of using the proposed target-aware attention to capture diverse user interests.

\section{Conclusions}

In this paper, we have developed a novel target attentive graph neural network model for session-based recommendation. By incorporating graph modeling and a target-aware attention module, the proposed TAGNN jointly considers user interests given a certain target item as well as complex item transitions in sessions. We have conducted thorough empirical evaluation to investigate TAGNN. Extensive experiments on real-world benchmark datasets demonstrate the effectiveness of our model.

\begin{acks}
This work is jointly supported by National Key Research and Development Program (2018YFB1402600, 2016YFB1001000) and National Natural Science Foundation of China (U19B2038, 61772528).
\end{acks}

\bibliographystyle{ACM-Reference-Format}
\bibliography{sigir}

\end{document}